\documentclass[twocolumn,showpacs,superscriptaddress,aps,prb]{revtex4-1}
\usepackage{latexsym}
\usepackage{amssymb}
\usepackage[pdftex, colorlinks,linkcolor=green,urlcolor=blue, anchorcolor=blue,citecolor=green]{hyperref}
\usepackage{xcolor}
\usepackage{natbib}

\usepackage{graphicx}
\newcommand{\bq}{\begin{equation}}
\newcommand{\eq}{\end{equation}}
\newcommand{\bn}{\begin{eqnarray}}
\newcommand{\en}{\end{eqnarray}}
\begin{document}
\title{ On the competition between the Kondo effect and the exchange interaction in a parallel double quantum dot system}

\author { Guo-Hui Ding}
\affiliation{Key Laboratory of Artificial Structures and Quantum Control (Ministry of Education), Department of Physics and Astronomy, Shanghai Jiao Tong University, 800 Dongchuan Road, Shanghai 200240, China}
\author {Fei Ye}
\affiliation{ Department of Physics, South University of Science and Technology of China, Shenzhen 518055, China}
\author { Xiaoqun Wang}
\affiliation{Key Laboratory of Artificial Structures and Quantum Control (Ministry of Education), Department of Physics and Astronomy, Shanghai Jiao Tong University, 800 Dongchuan Road, Shanghai 200240, China}

\date{\today }

\begin{abstract}
We study the competition between the Kondo effect and the exchange interaction in the parallel double quantum dot(DQDs) system within an effective action field theory. The strong on-site Coulomb interactions in DQDs are treated by using the Hubbard-Stratonovich transformation and the introduction of scalar potential fields. We show that a self-consistent perturbation approach, which takes into account the statistical properties of the potential fields acting on the electrons in DQDs,  describes well the crossover from the Kondo regime to the spin-singlet state in this system. The linear conductance and the intradot/interdot spin excitation spectra of this system are obtained.

\end{abstract}
 \maketitle

\newpage
\section{Introduction}
The quantum transport through semiconductor quantum dots \cite{vanderWiel2002,Hanson2007} has
attracted great research attentions in recent decades, because of many interesting physical phenomena,  e.g., the
Coulomb blockade effect,\cite{Houten1992} the Kondo effect,\cite{GoldhaberGordon1998,Cronernwett1998,vanderWiel2000} non-Fermi liquid states,\cite{Potok2007} etc., observed in these systems and also their possible applications in building new generation nanoscale electronic devices.  The strong Coulomb interaction of electrons in nanoscale structures poses a great challenge for the theoretically treatment. A variety of theoretical methods have been applied to study the transport properties of QD systems both in the linear response region and the nonlinear out-of equilibrium case,  including the numerical renormalization group (NRG) method,\cite{Bulla2008} impurity-DMFT,\cite{Mitchell2015} the functional RG approach,\cite{Metzner2012} the noncrossing approximation, \cite{Wingreen1994} the fluctuation exchange approximation (FLEX), \cite{Horvath2011,Bickers1989, Bickers1991} and continuous-time Monte Carlo methods,\cite{Gull2011} etc. These methods all have achieved some remarkable successes but also with different limitations. For instance, the NRG method excellently describes the Kondo effect at low temperature, which is a paradigm strong correlated effect arising from the spin exchange interaction between  electrons localized in QDs and the conduction electrons in the metallic leads, but it is difficult to study out-of equilibrium transport properties within this method.

In this work we will study the electron transport through a parallel DQD system by using a self-consistent perturbation method within the two-particle-irreducible (2PI) effective action field theory/Kadanoff-Baym's $\Phi$-functional theory.\cite{Baym1962,Kadanoff1961,Cornwall1974} This method has recently been applied to study the transient current behaviors of the single Anderson impurity model.\cite{Sexty2011,Sebastian2016} We will show that the effective action field theory based on the Hubbard-Stratonovich transformation  does capture the main features of strong correlation effects in DQDs, and correctly describes the competition between the Kondo correlation and the exchange interaction in the ground state of this system.  One benefit of this method is that it can be  readily  generalized to high dimensional correlated electron systems. For parallel DQDs connected to the same source and drain leads, the quantum phase transitions in this system have been investigated in some previous works.\cite{Zitko2007,Zitko2012,Ding2009,Ding2005,Horvath2011,Wong2012}  It is known that without interdot tunnel coupling a ferromagnetic spin exchange interaction between DQDs meditated by tunneling to leads is generated,  and leads to the underscreened Kondo effect and singular Fermi liquid behaviors\cite{Posazhennikova2007} at low temperature. With increasing the interdot tunnel coupling, a quantum phase transition from the Kondo regime to a spin singlet state is found in the ground state. Both of the strong Coulomb interaction and the Fano interference effect greatly influence the linear conductance of this system.\cite{Ding2005}
In the present work, we will consider a parallel DQD system in another configuration (shown schematically in Fig. 1):  A DQD system with interdot tunneling coupling and each of two QDs connected to its own source and drain leads. This kind of DQD systems have also attracted a lot of research interests due to the possible existence of the orbital Kondo effect,\cite{Pohjola2001,Kubo2008} the  complex quantum phase transitions\cite{Zarand2006,Galpin2005}  and the Coulomb drag effect\cite{Keller2016,Sanchez2010,Kaasbjerg2016} when DQDs have solely the spin exchange interaction or the capacity coupling. It was  pointed out that the quantum phase transition in this DQD system is unstable to the charge transfer between two QDs.\cite{Zarand2006}

\begin{figure}[htp]
\includegraphics[width=0.8\columnwidth,height=2.0in,angle=0]{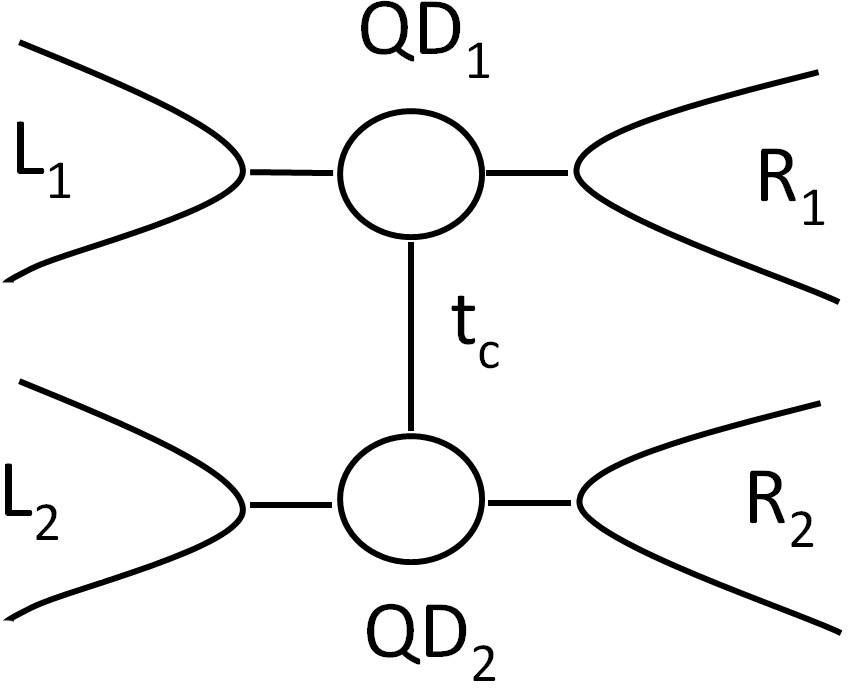}
\caption{ The schematic diagram of the double quantum dot system.}
\end{figure}

\section{ Effective action and the Method}
We describe the electron transport through the DQDs  by a two-impurity Anderson model. Within the path-integral formulation on the closed-time Keldysh contour, the action of this model is given by
\bn
\Gamma&=&\int_C dt \left \{ \sum_{i,\sigma} \left [d^\dagger_{i\sigma}(i{\frac {\partial}{\partial t}}-\epsilon_{i\sigma})d_{i\sigma}-U  n_{i\uparrow}  n_{i\downarrow}\right ]\right.\nonumber\\
&+&\sum_{k,i,\eta_i,\sigma}\left [c^\dagger_{k\eta_i\sigma}(i{\frac {\partial}{\partial t}}-\epsilon_{k\eta_i})c_{k\eta_i\sigma}
-( v_{k\eta_i} c^\dagger_{k\eta_i\sigma}d_{i\sigma}+{\rm H.c.})\right ]
\nonumber\\
&-&\left. t_c\sum_\sigma (d^\dagger_{1\sigma}d_{2\sigma}+d^\dagger_{2\sigma}d_{1\sigma})
 \right \}\;,
\en
where $i\in \{1,2\}$ denotes two different QDs, $\sigma\in \{\uparrow, \downarrow\}$ is the electron spin index, and $\eta_i$ denotes the left and right leads coupled to the $i$th-QD ($\eta_i=L_i, R_i$). $d_{i\sigma}$($d^\dagger_{i\sigma}$) and $c_{k\eta_i\sigma}$($c^\dagger_{k\eta_i\sigma}$) are the Grassmann variables for the electron operators of QDs and leads, respectively.  $v_{k\eta_i}$ describes the tunneling matrix element between the lead and the QD.  $U$ is on-site Coulomb interaction strength. $t_c$ is the interdot tunnel coupling. By integrating out the Grassmann variables of the leads and making the Hubbard-Stratonovich transformation for the Coulomb interaction term, we can obtain an effective action for the QD variables
\bn
\Gamma_{eff}&=&\int_C dt \int _C dt' \sum_{i,\sigma} d^\dagger_{i\sigma}(t)\left [(i{\frac {\partial}{\partial t}}-\epsilon_{i\sigma}-\phi_{i\sigma})\delta_c(t-t')\right.
\nonumber\\
&-&\left.\Sigma^{(0)}_i(t,t')\right ]d_{i\sigma}(t')+\int_C dt \left [{\frac {1}{U}} \sum_i \phi_{i\uparrow}\phi_{i\downarrow} \right.
\nonumber\\
&-& \left. t_c\sum_\sigma (d^\dagger_{1\sigma}d_{2\sigma}+d^\dagger_{2\sigma}d_{1\sigma})\right ]\;,
\en
Here $\Sigma^{(0)}_i(t,t')\equiv \sum_{k\eta_i} |v_{k\eta_i}|^2 g_{k\eta_i}(t,t')$ is a self-energy term of the $i$th QD induced by  the tunnel coupling with the leads, with $g_{k\eta_i}(t,t')$ being the bare GF of the lead $\eta_i$. The scalar field $\phi_{i\sigma}$, which represents the fluctuating potential acting on  electrons in the QD,  is introduced by the Hubbard-Stratonovich transformation. One can replace the field $\phi_{i\sigma}$ in the effective action as the sum of its mean value and the fluctuation part: $\phi_{i\sigma}=\langle \phi_{i\sigma} \rangle
 +\delta \phi_{i\sigma}$, where  the time-dependent scalar field  $\delta \phi_{i\sigma}$ is introduced to describe the fluctuation of the potential field.  In order to simplify the notation, we introduce  a two component Fermi field $d_\sigma\equiv\left (\begin {array}{c}
d_{1\sigma} \\
d_{2\sigma}
\end{array}
\right )$
and a four component scalar field $\delta\phi \equiv\left (\begin {array}{c}
\delta\phi_{1\uparrow} \\
\delta\phi_{1\downarrow} \\
\delta\phi_{2\uparrow} \\
\delta\phi_{2\uparrow} \\
\end{array}
\right )$. Then the effective action can be written as
\bn
&&\Gamma_{eff}=\int_C dt\int_C dt' \left [ \sum_\sigma d^\dagger_\sigma(t) G^{-1}_{0\sigma}(t,t')d_\sigma(t')\right.\nonumber\\
&&\left.+{\frac{1}{2}}\delta\phi^T(t)D^{-1}_{{0}}(t,t')\delta\phi(t')\right ]+ \frac {1}{U} \int_C dt \sum_i \langle\phi_{i\uparrow}\rangle\langle\phi_{i\downarrow}\rangle \nonumber\\
&&+I_{int}(d^\dagger_{i\sigma},d_{i\sigma},\delta\phi_{i\sigma})\;.
\en
 where the inverse of the bare GF for the Fermi field is given by:\\
$G^{-1}_{0\sigma}(t,t')=\left [ i\frac {\partial}{\partial t}- \left (\begin {array}{cc}
\epsilon_{1\sigma}+\langle\phi_{1\sigma}\rangle  & t_c\\
t_c  & \epsilon_{2\sigma}+\langle\phi_{2\sigma}\rangle
\end{array}
\right )\delta_c(t-t')\right.
\\
-\left.\left (\begin {array}{cc}
\Sigma^{(0)}_1(t,t') & 0\\
0  &  \Sigma^{(0)}_2(t,t')
\end{array}
\right ) \right ]\;,
$
 and that of the bare GF of the scalar field is:
$D^{-1}_{0}(t,t')={\frac {1}{U}}\left (\begin {array}{cccc}
0 & 1 & 0 & 0\\
1 & 0 & 0 & 0 \\
0 & 0 & 0 & 1\\
0 & 0 & 1 & 0 \\
\end{array}
\right )\delta_c(t-t')$.
The last term in Eq.(3) is the interaction action term $I_{int} $ given by
\bq
I_{int}=-\int_C dt \sum_{i,\sigma} \delta\phi_{i\sigma} (d^\dagger_{i\sigma}d_{i\sigma}-\langle n_{i\sigma}\rangle)\;.
\eq
where the expectation value of the spin-resolved dot occupation number $\langle n_{i\sigma}\rangle$ is related to the mean value of the potential field: $\langle \phi_{i\bar\sigma}\rangle=U \langle n_{i\sigma}\rangle$.

\begin{figure}[htp]
\includegraphics[width=0.9\columnwidth,height=2.0in,angle=0]{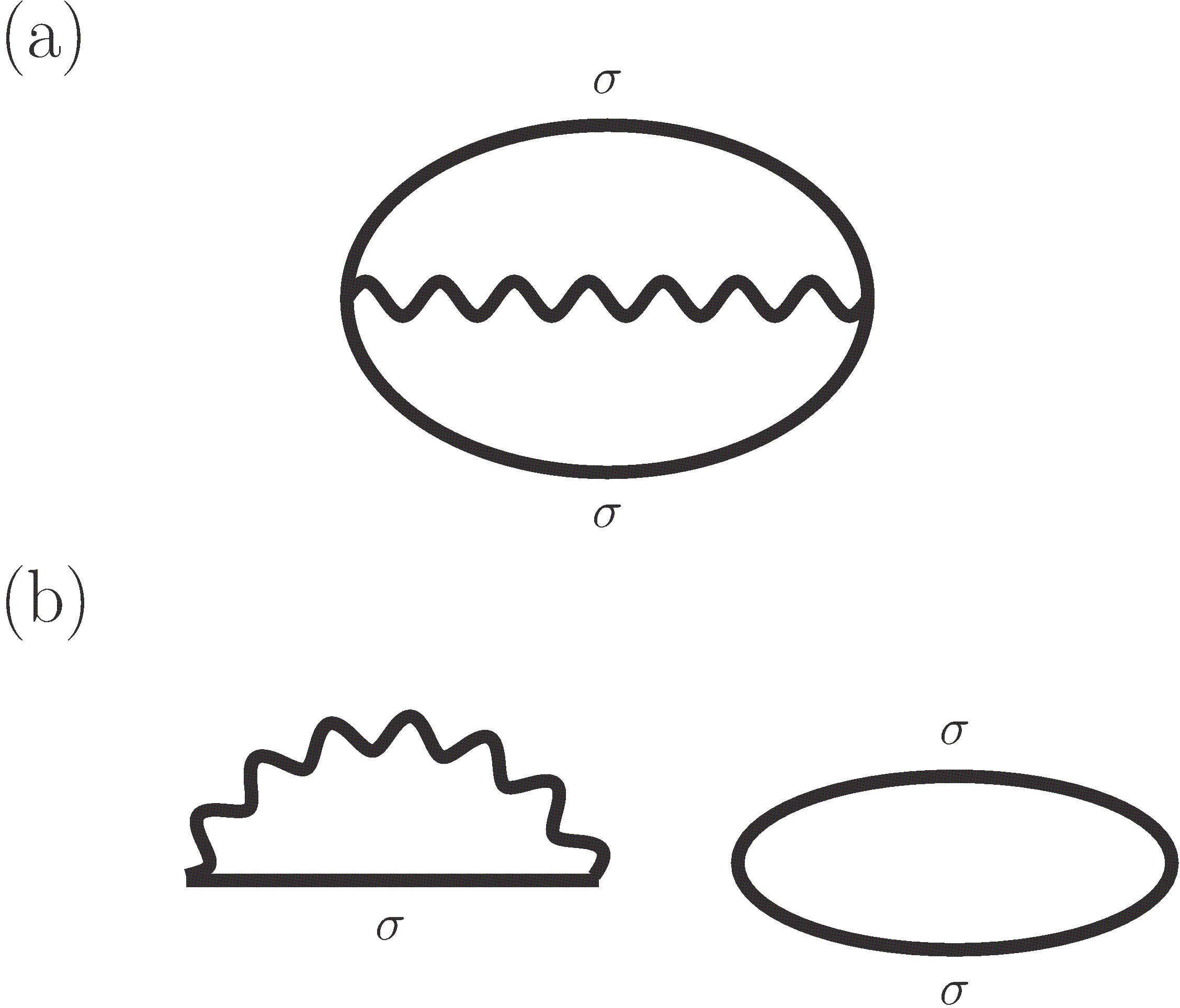}
\caption{(a) The Feynman diagram for the second-order perturbation term in the effective action; (b)The Feynman diagrams for the self energies of the Fermi field and the scalar potential field. }
\end{figure}

\begin{figure}[htp]
\includegraphics[width=\columnwidth,height=4.0in,angle=0]{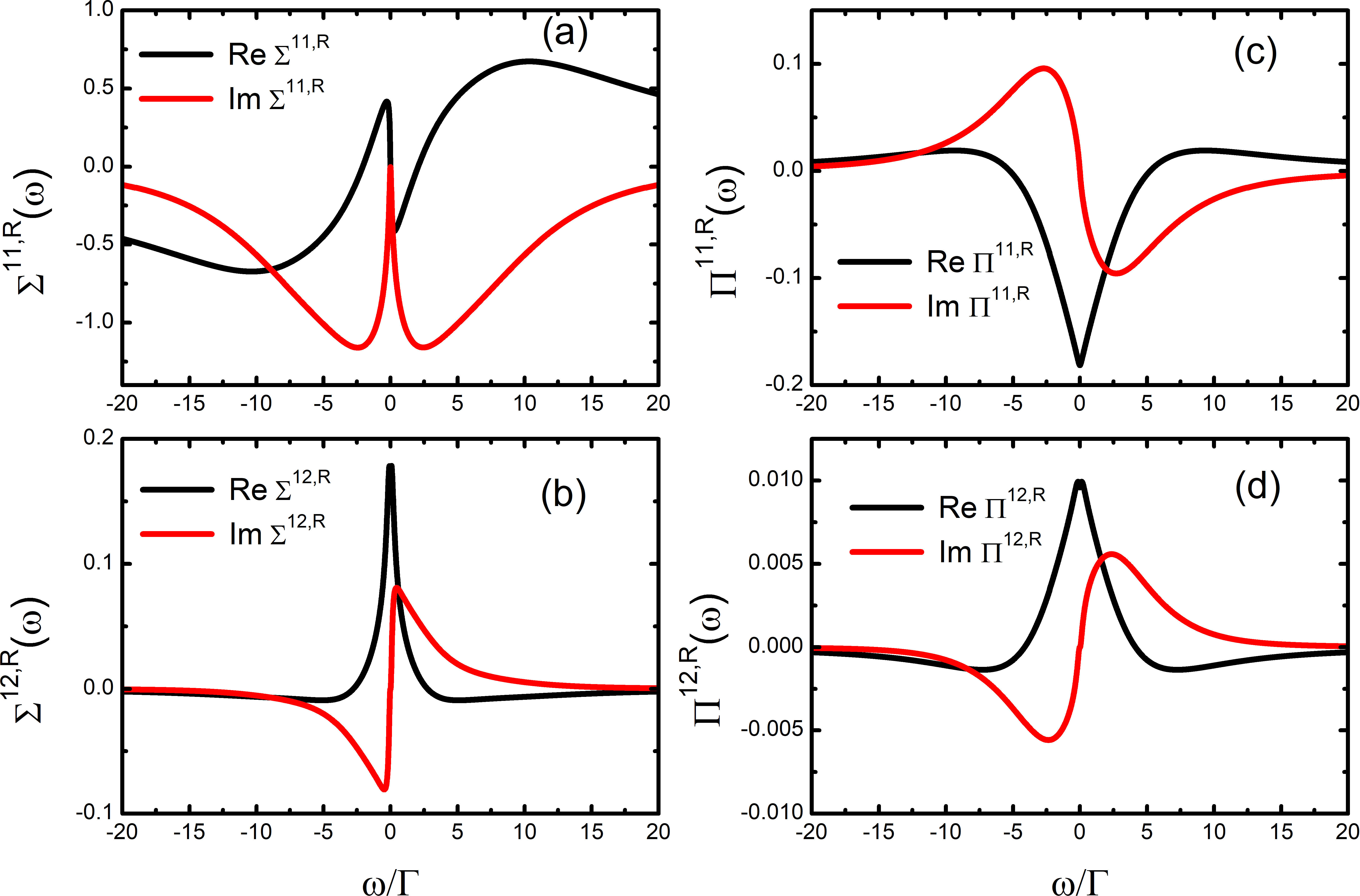}
\caption{ The  self-energies for DQDs with interdot tunneling coupling. (a) and (b) The intradot and interdot self-energies  for the Fermi field, respectively; (c) and (d) The intradot and interdot self-energies of the scalar potential field.  The parameters for the model: $\epsilon_1=\epsilon_2=-U/2$, $U/\Gamma=5.0$, and  $t_c=0.5$.}
\end{figure}

Now we will treat the interaction term by using a self-consistent perturbation method. Within the 2PI effective action theory, one introduces self-energies  $\Sigma_\sigma$ and $\Pi$ for the Fermi field and the scalar potential field, respectively. Then the inverses of the full GFs can be written as: $G^{-1}_\sigma(t,t')= G^{-1}_{0\sigma}(t,t')-\Sigma_\sigma(t,t')$; and $D^{-1}(t,t')= D^{-1}_{0}(t,t')-\Pi(t,t')$. Here the self-energy $\Sigma_\sigma$  is in a $2\times 2$ matrix form: $\Sigma_\sigma=\left (\begin{array}{cc}
\Sigma^{11}_\sigma  & \Sigma^{12}_\sigma\\
\Sigma^{21}_\sigma  & \Sigma^{22}_\sigma
\end{array} \right ) $,
and the self-energy $\Pi$ is a $4\times 4$ matrix:   $\Pi=\left (\begin{array}{cc}
\Pi^{11}_{\sigma\sigma'}  & \Pi^{12}_{\sigma\sigma'}\\
\Pi^{21}_{\sigma\sigma'}  & \Pi^{22}_{\sigma\sigma'}
\end{array} \right ) $, with spin indexs $\sigma, \sigma'\in\{\uparrow, \downarrow\}$.
Following a standard procedure\cite{Sexty2011,Sebastian2016}, one can obtain the 2PI effective action  as
\bn
&&\Gamma[G,D,\langle\phi\rangle]=-i {\rm Tr}[\ln G^{-1}+G^{-1}_{0}G]+\frac {1}{U} {\rm Tr}(\langle\phi_{i\uparrow}\rangle\langle\phi_{i\downarrow}\rangle)\nonumber\\
&&+\frac{i}{2} {\rm Tr}[\ln D^{-1}+D^{-1}_{0}D]+\Gamma_2[G,D]+const.\;,
\en
where the trace $\rm Tr$ means  sums over the necessary spin index or dot index,  and also the integral on the closed-time Keldysh contour. $\Gamma_2[G,D]$ contains all closed 2PI diagrams obtained from the the expectation value functional of the interaction action term: $\exp({i\Gamma_2})=\langle \exp({i I_{int}})\rangle$.  The lowest-order contribution to the effective action comes from the connected Feynman diagram of the second-order term  $\Gamma_2\approx  {\frac{i}{2}}\langle   I_{int}^2\rangle_c$,  as shown schematically in Fig. 2(a). It reads
\bn
\Gamma_2[G,D]&=&-\frac{1}{2}\int_C dt\int_C dt' \sum_{i,j,\sigma} G_{ij,\sigma}(t,t')G_{ji,\sigma}(t',t)
\nonumber\\
&&D_{ij,\sigma\sigma}(t,t')\;.
\en
Then the stationarity conditions of the effective action
 lead to a set of self-consistent equations.  For instance, the condition $\delta \Gamma/\delta \langle\phi_{i\bar\sigma}\rangle=0$, gives the self-consistent equation of the potential field: $\langle \phi_{i\sigma}\rangle=-i U G_{ii,\bar\sigma}(t,t^+)$. The self-consistent equation for the electron self-energy  is
\bq
\Sigma^{ij}_\sigma(t,t')=-i \frac{\delta\Gamma_2[G,D]}{\delta G_{ji,\sigma}(t',t)}=iG_{ij,\sigma}(t,t')D_{ij,\sigma\sigma}(t,t')\;,
\eq
and the self-energy of the scalar potential field
\bq
\Pi^{ij}_{\sigma\sigma}(t,t')=2i \frac{\delta\Gamma_2[G,D]}{\delta D_{ji,\sigma\sigma}(t',t)}=-iG_{ij,\sigma}(t,t')G_{ji,\sigma}(t',t)
\eq
and $\Pi^{ij}_{\sigma\bar\sigma}(t,t')=0$. The corresponding Feynman diagrams of the self-energies are plotted in Fig. 2(b).

\begin{figure}[htp]
\includegraphics[width=\columnwidth,height=4.0in,angle=0]{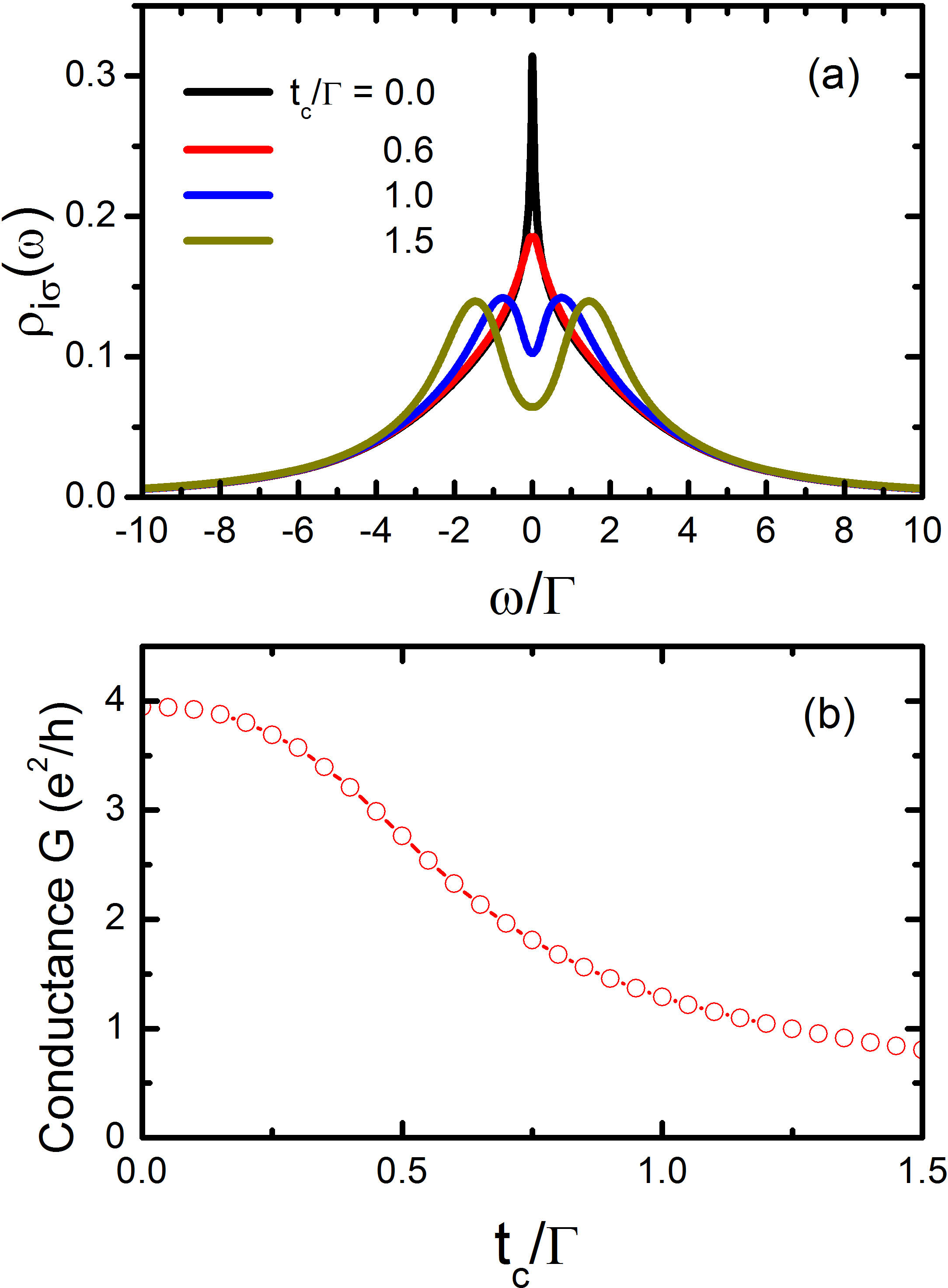}
\caption{ (a) The density of state of the quantum dots with different interdot tunnel coupling $t_c$;
(b) The linear conductance $G$ vs. the interdot tunnel coupling $t_c$.  }
\end{figure}

\begin{figure}[htp]
\includegraphics[width=\columnwidth,height=3.5in,angle=0]{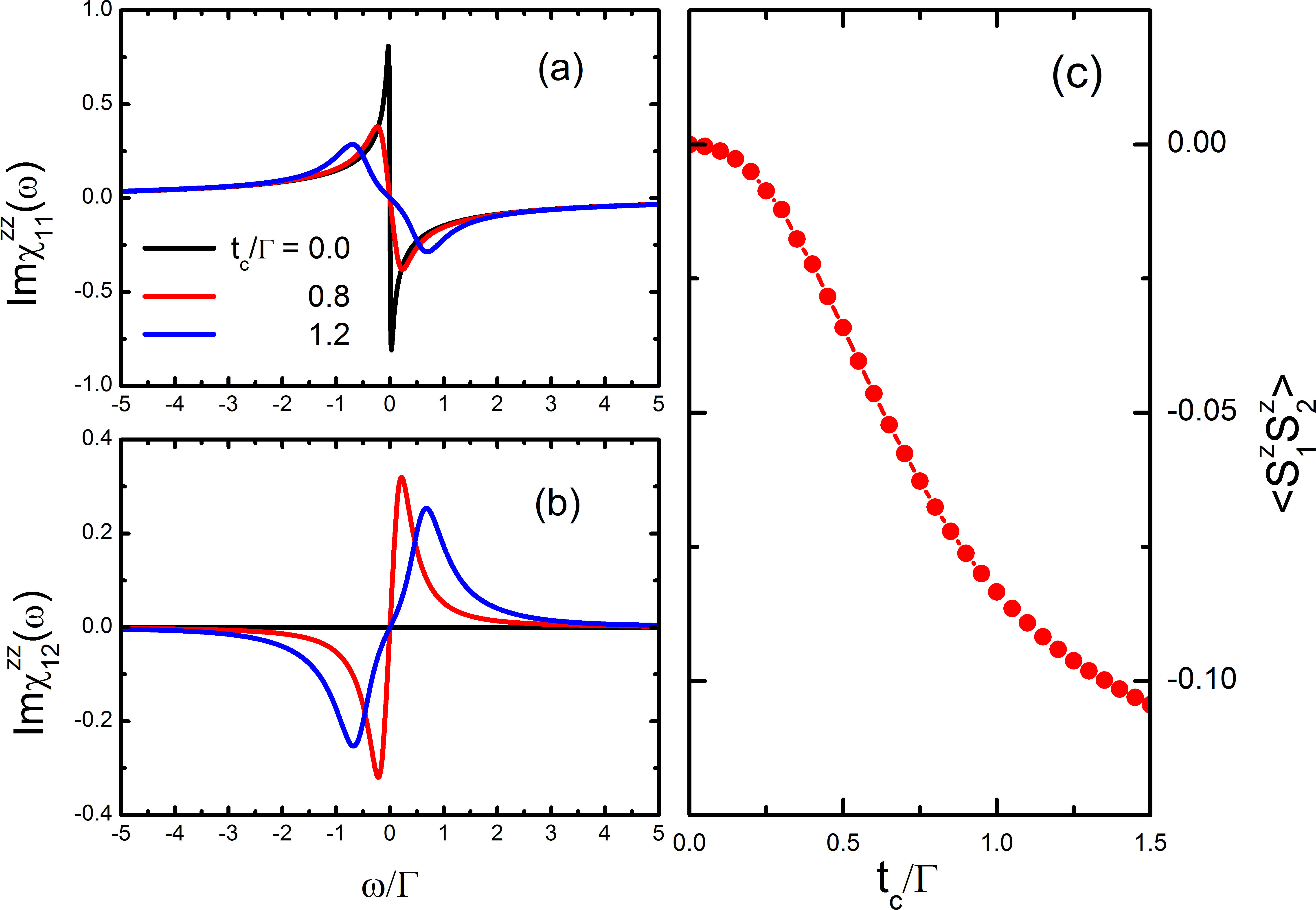}
\caption{(a) The imaginary part of the intradot spin correlation function; (b) The imaginary part of the interdot spin correlation function; (c) The static spin correlation $\langle S^z_1S^z_2\rangle$ vs. the interdot tunnel coupling $t_c$.  }
\end{figure}

\section{Numerical results and discussion}

We first obtain the self-energies and also the dressed GFs for the Fermi field and the scalar potential field by solving the self-consistent equations  numerically with the fast Fourier transform (FFT) method. A DQD system  at zero temperature with degenerate dot levels and symmetric couplings to the left and the right leads is considered. The hybridization strength between the dots and the leads is denoted as $\Gamma$.  In our calculation we take the model parameters: $\Gamma=1$, $U=5.0$, $\epsilon_1=\epsilon_2=-U/2$,  thereby the system has the particle-hole symmetry.  In the equilibrium case, the chemical potentials of leads are assumed as $\mu_{L_i}=\mu_{R_i}=0$ .

In Fig. 3 we plot the self-energies of the retarded GFs for the Fermi field and the potential field.  DQDs with a particular  value of interdot tunnel coupling $t_c$  are considered. It is noted that the self-energy terms contain both the intradot parts and interdot parts, and they exhibit quite different structures in their frequency dependent properties. It is interesting to notice that the real part of the intradot self-energy for the scalar potential field shows a broad dip structure in the low frequency region, whereas the real part of the interdot self-energy of the scalar potential field shows a broad peak structure. This broad peak and dip structures represent important statistical properties of the scalar potential fields emerged from  the strong on-site Coulomb interaction.

The local density of states (LDOS) in each dot can be easily obtained from the dressed GF: $\rho_{i\sigma}(\omega)=-{\rm Im}G^r_{ii,\sigma}(\omega)/\pi$. In Fig. 4(a) the LDOS $\rho_{i\sigma}(\omega)$ for DQDs with different interdot tunnel coupling $t_c$ are plotted. In the absence of interdot tunneling ($t_c=0$), the LDOS for each of two QDs exhibits a sharp peak in zero frequency regime because of the Kondo effect. With increasing the interdot tunnel coupling $t_c$, the Kondo effect is gradually suppressed.  Beyond some critical value of $t_c$, a two side-peak structure of LDOS is observed  in the  low frequency region, which can be regarded as the formation of a spin singlet state between electrons located in DQDs.

When a bias voltage is symmetrically applied to the leads (with $\mu_{L_i}=-\mu_{R_i}=\Delta\mu/2$), the total linear conductance $G$ of this system can be calculated by using the formula: $G={\frac{e^2}{h}}\sum_\sigma{\rm Tr }[-\Gamma {\rm Im} G_\sigma(\omega=0)]$. In Fig. 4(b) one can see that in the absence of interdot tunneling,  the conductance $G$ reaches the unitary limit  $G=4e^2/h$ for this two independent channel system. We find that the conductance decreases continuously with the increasing of the interdot tunnel coupling $t_c$. Thereby the transition of the ground state of this DQDs from the Kondo regime to the spin singlet state will be a smooth crossover and this system doesn't display any abrupt quantum phase transitions, which is in agreement with the conclusion obtained from the conformal field theory and RG approach.\cite{Zarand2006} The reason might be the presence of effective charge transfer processes between  two different  electron transport channels in the effective Hamiltonian  for the low-energy excitations.\cite{Zarand2006}

In order to ensure that a spin singlet is formed in the ground state for the system with large interdot tunnel coupling, we examine both the intradot and interdot spin correlation functions in DQDs. The spin correlation functions will be defined as  $\chi^{zz}_{ij}(t,t')\equiv -i\langle T_c S^z_i(t)S^z_j(t')\rangle$, where the $z$-component of the spin operator in the QD $S^z_i=(n_{i\uparrow}-n_{i\downarrow})/2$.   To obtain correlation functions of spin operators, one has to calculate various correlation functions of spin-resolved number operators  $\langle T_c\delta n_\sigma(t)\delta n_{\sigma'}(t')\rangle$.  We  make the RPA approximation in our calculation of  correlation functions $\langle T_c\delta n_\sigma(t)\delta n_{\sigma'}(t')\rangle$ by using a functional derivative method,\cite{Ding2013} and here the RPA approximation is equivalent to only consider  vertex correction terms contributed from the functional derivatives of the Hartree term $U\langle n_{\bar\sigma}\rangle$ in the self-energy of GFs with respect to external potential fields.  The numerical results for the intradot and the interdot spin dynamics are plotted in Fig.5 (a) and (b).  Without the interdot tunnel coupling ($t_c=0$),  the imaginary part of the intradot  retarded spin correlation function ${\rm Im}\chi^{zz,r}_{11}(\omega)$ has a sharp peak in the energy scale of the Kondo temperature $T_K$, which is a characteristic of the spin excitation spectrum in the Kondo regime.  Since two QDs are decoupled in this case, the interdot spin correlation function ${\rm Im}\chi^{zz,r}_{12}(\omega)$  is exactly zero as shown in Fig. 5 (b). For the system with interdot tunneling ($t_c\neq 0$), the intradot spin excitation spectrum ${\rm Im}\chi^{zz,r}_{11}(\omega)$  becomes much more broad and shifts to the higher frequency region.  The interdot spin excitation spectrum ${\rm Im}\chi^{zz,r}_{12}(\omega)$ also becomes significantly large.

The static spin correlation between QDs $\langle S^z_1S^z_2\rangle$ can be obtained according to the fluctuation-dissipation theorem: $\langle S^z_1S^z_2\rangle=-\int^\infty_0 d\omega {\rm Im}\chi ^{zz,r}_{12}(\omega)/\pi$. In Fig. 5(c) the static spin correlation $\langle S^z_1S^z_2\rangle$ vs. the interdot tunnel coupling $t_c$ is plotted.   $\langle S^z_1S^z_2\rangle$ always has a negative value as long as $t_c\neq 0$, indicating an antiparallel configuration of the electron spins in DQDs is favored.  The  value of static spin correlation decreases significant when  $t_c$ increases, e.g.,  $\langle S^z_1S^z_2\rangle$ is lesser than $-0.1$ as $t_c>1.5$. We can regard it as an evidence of a spin singlet like state  formed in the many body ground state of the DQDs. Since it is well known that for an isolate spin singlet the values of the static spin correlation $\langle S^z_1S^z_2\rangle=-0.25$, and in this DQD system both of charge fluctuations and many body correlation effects will influence the static spin correlation.

\section{Summary}

In summary,  we have studied the electron transport through a parallel DQD system with interdot tunnel coupling  and strong  on-site Coulomb interactions.  The introduction of scalar potential fields acting on  electrons in QDs by using the Hubbard-Stratonovich transformation and the quantification of the statistical properties of fluctuating potential fields are essential to take into account the charge and spin fluctuations in this system. We describe the competition between the Kondo effect and the exchange interaction by a self-consistent perturbation theory within the 2PI effective action formulation.  We find that there is  a continuous crossover from the Kondo regime to the spin singlet regime in the ground state of this DQD system as  the interdot tunnel coupling increases.  One can expect that this self-consistent perturbation theory on the Keldysh contour presented in this work is not limited to be useful in the study of quantum transport through  zero-dimensional QDs, but also has important applications in the study of quantum phase transitions in  higher dimensional strong-correlated electron systems.

\begin{acknowledgments}
{This work was supported by the National Natural Science Foundation of China (Grant No: 11674223).}
\end{acknowledgments}

\bibliography{RefsDQD}

\begin{thebibliography}{34}%
\makeatletter
\providecommand \@ifxundefined [1]{%
 \@ifx{#1\undefined}
}%
\providecommand \@ifnum [1]{%
 \ifnum #1\expandafter \@firstoftwo
 \else \expandafter \@secondoftwo
 \fi
}%
\providecommand \@ifx [1]{%
 \ifx #1\expandafter \@firstoftwo
 \else \expandafter \@secondoftwo
 \fi
}%
\providecommand \natexlab [1]{#1}%
\providecommand \enquote  [1]{``#1''}%
\providecommand \bibnamefont  [1]{#1}%
\providecommand \bibfnamefont [1]{#1}%
\providecommand \citenamefont [1]{#1}%
\providecommand \href@noop [0]{\@secondoftwo}%
\providecommand \href [0]{\begingroup \@sanitize@url \@href}%
\providecommand \@href[1]{\@@startlink{#1}\@@href}%
\providecommand \@@href[1]{\endgroup#1\@@endlink}%
\providecommand \@sanitize@url [0]{\catcode `\\12\catcode `\$12\catcode
  `\&12\catcode `\#12\catcode `\^12\catcode `\_12\catcode `\%12\relax}%
\providecommand \@@startlink[1]{}%
\providecommand \@@endlink[0]{}%
\providecommand \url  [0]{\begingroup\@sanitize@url \@url }%
\providecommand \@url [1]{\endgroup\@href {#1}{\urlprefix }}%
\providecommand \urlprefix  [0]{URL }%
\providecommand \Eprint [0]{\href }%
\providecommand \doibase [0]{http://dx.doi.org/}%
\providecommand \selectlanguage [0]{\@gobble}%
\providecommand \bibinfo  [0]{\@secondoftwo}%
\providecommand \bibfield  [0]{\@secondoftwo}%
\providecommand \translation [1]{[#1]}%
\providecommand \BibitemOpen [0]{}%
\providecommand \bibitemStop [0]{}%
\providecommand \bibitemNoStop [0]{.\EOS\space}%
\providecommand \EOS [0]{\spacefactor3000\relax}%
\providecommand \BibitemShut  [1]{\csname bibitem#1\endcsname}%
\let\auto@bib@innerbib\@empty
\bibitem [{\citenamefont {van~der Wiel}\ \emph {et~al.}(2002)\citenamefont
  {van~der Wiel}, \citenamefont {De~Franceschi}, \citenamefont {Elzerman},
  \citenamefont {Fujisawa}, \citenamefont {Tarucha},\ and\ \citenamefont
  {Kouwenhoven}}]{vanderWiel2002}%
  \BibitemOpen
  \bibfield  {author} {\bibinfo {author} {\bibfnamefont {W.~G.}\ \bibnamefont
  {van~der Wiel}}, \bibinfo {author} {\bibfnamefont {S.}~\bibnamefont
  {De~Franceschi}}, \bibinfo {author} {\bibfnamefont {J.~M.}\ \bibnamefont
  {Elzerman}}, \bibinfo {author} {\bibfnamefont {T.}~\bibnamefont {Fujisawa}},
  \bibinfo {author} {\bibfnamefont {S.}~\bibnamefont {Tarucha}}, \ and\
  \bibinfo {author} {\bibfnamefont {L.~P.}\ \bibnamefont {Kouwenhoven}},\
  }\href {\doibase 10.1103/RevModPhys.75.1} {\bibfield  {journal} {\bibinfo
  {journal} {Rev. Mod. Phys.}\ }\textbf {\bibinfo {volume} {75}},\ \bibinfo
  {pages} {1} (\bibinfo {year} {2002})}\BibitemShut {NoStop}%
\bibitem [{\citenamefont {Hanson}\ \emph {et~al.}(2007)\citenamefont {Hanson},
  \citenamefont {Kouwenhoven}, \citenamefont {Petta}, \citenamefont {Tarucha},\
  and\ \citenamefont {Vandersypen}}]{Hanson2007}%
  \BibitemOpen
  \bibfield  {author} {\bibinfo {author} {\bibfnamefont {R.}~\bibnamefont
  {Hanson}}, \bibinfo {author} {\bibfnamefont {L.~P.}\ \bibnamefont
  {Kouwenhoven}}, \bibinfo {author} {\bibfnamefont {J.~R.}\ \bibnamefont
  {Petta}}, \bibinfo {author} {\bibfnamefont {S.}~\bibnamefont {Tarucha}}, \
  and\ \bibinfo {author} {\bibfnamefont {L.~M.~K.}\ \bibnamefont
  {Vandersypen}},\ }\href {\doibase 10.1103/RevModPhys.79.1217} {\bibfield
  {journal} {\bibinfo  {journal} {Rev. Mod. Phys.}\ }\textbf {\bibinfo {volume}
  {79}},\ \bibinfo {pages} {1217} (\bibinfo {year} {2007})}\BibitemShut
  {NoStop}%
\bibitem [{\citenamefont {van Houten}\ \emph {et~al.}(1992)\citenamefont {van
  Houten}, \citenamefont {Beenakker},\ and\ \citenamefont
  {Staring}}]{Houten1992}%
  \BibitemOpen
  \bibfield  {author} {\bibinfo {author} {\bibfnamefont {H.}~\bibnamefont {van
  Houten}}, \bibinfo {author} {\bibfnamefont {C.~W.~J.}\ \bibnamefont
  {Beenakker}}, \ and\ \bibinfo {author} {\bibfnamefont {A.~A.~M.}\
  \bibnamefont {Staring}},\ }\href@noop {} {\emph {\bibinfo {title} {in Single
  Charge Tunneling}}}\ (\bibinfo  {publisher} {edited by H. Grabert and M. H.
  Devoret, NATO Advanced Studies Institutes, Series B: Physics, Vol. 294
  Plenum, New York},\ \bibinfo {year} {1992})\BibitemShut {NoStop}%
\bibitem [{\citenamefont {Goldhaber-Gordon}\ \emph {et~al.}(1998)\citenamefont
  {Goldhaber-Gordon}, \citenamefont {Shtrikman}, \citenamefont {Mahalu},
  \citenamefont {Abusch-Magder}, \citenamefont {Meirav},\ and\ \citenamefont
  {Kastner}}]{GoldhaberGordon1998}%
  \BibitemOpen
  \bibfield  {author} {\bibinfo {author} {\bibfnamefont {D.}~\bibnamefont
  {Goldhaber-Gordon}}, \bibinfo {author} {\bibfnamefont {H.}~\bibnamefont
  {Shtrikman}}, \bibinfo {author} {\bibfnamefont {D.}~\bibnamefont {Mahalu}},
  \bibinfo {author} {\bibfnamefont {D.}~\bibnamefont {Abusch-Magder}}, \bibinfo
  {author} {\bibfnamefont {U.}~\bibnamefont {Meirav}}, \ and\ \bibinfo {author}
  {\bibfnamefont {M.~A.}\ \bibnamefont {Kastner}},\ }\href
  {https://www.nature.com/articles/34373} {\bibfield  {journal} {\bibinfo
  {journal} {Nature}\ }\textbf {\bibinfo {volume} {391}},\ \bibinfo {pages}
  {156} (\bibinfo {year} {1998})}\BibitemShut {NoStop}%
\bibitem [{\citenamefont {Cronenwett}\ \emph {et~al.}(1998)\citenamefont
  {Cronenwett}, \citenamefont {Oosterkamp},\ and\ \citenamefont
  {Kouwenhoven}}]{Cronernwett1998}%
  \BibitemOpen
  \bibfield  {author} {\bibinfo {author} {\bibfnamefont {S.~M.}\ \bibnamefont
  {Cronenwett}}, \bibinfo {author} {\bibfnamefont {T.~H.}\ \bibnamefont
  {Oosterkamp}}, \ and\ \bibinfo {author} {\bibfnamefont {L.~P.}\ \bibnamefont
  {Kouwenhoven}},\ }\href
  {http://science.sciencemag.org/content/sci/281/5376/540.full.pdf} {\bibfield
  {journal} {\bibinfo  {journal} {Science}\ }\textbf {\bibinfo {volume}
  {281}},\ \bibinfo {pages} {540} (\bibinfo {year} {1998})}\BibitemShut
  {NoStop}%
\bibitem [{\citenamefont {van~der Wiel}\ \emph {et~al.}(2000)\citenamefont
  {van~der Wiel}, \citenamefont {Fujisawa}, \citenamefont {Elzerman},
  \citenamefont {Tarucha},\ and\ \citenamefont {Kouwenhoven}}]{vanderWiel2000}%
  \BibitemOpen
  \bibfield  {author} {\bibinfo {author} {\bibfnamefont {S.}~\bibnamefont
  {van~der Wiel}, \bibfnamefont {W.~G.and De~Franceschi}}, \bibinfo {author}
  {\bibfnamefont {T.}~\bibnamefont {Fujisawa}}, \bibinfo {author}
  {\bibfnamefont {J.~M.}\ \bibnamefont {Elzerman}}, \bibinfo {author}
  {\bibfnamefont {S.}~\bibnamefont {Tarucha}}, \ and\ \bibinfo {author}
  {\bibfnamefont {L.~P.}\ \bibnamefont {Kouwenhoven}},\ }\href
  {http://science.sciencemag.org/content/sci/289/5487/2105.full.pdf} {\bibfield
   {journal} {\bibinfo  {journal} {Science}\ }\textbf {\bibinfo {volume}
  {289}},\ \bibinfo {pages} {2105} (\bibinfo {year} {2000})}\BibitemShut
  {NoStop}%
\bibitem [{\citenamefont {Potok}\ \emph {et~al.}(2007)\citenamefont {Potok},
  \citenamefont {Rau}, \citenamefont {Shtrikman}, \citenamefont {Oreg},\ and\
  \citenamefont {Goldhaber-Gordon}}]{Potok2007}%
  \BibitemOpen
  \bibfield  {author} {\bibinfo {author} {\bibfnamefont {R.~M.}\ \bibnamefont
  {Potok}}, \bibinfo {author} {\bibfnamefont {I.~G.}\ \bibnamefont {Rau}},
  \bibinfo {author} {\bibfnamefont {H.}~\bibnamefont {Shtrikman}}, \bibinfo
  {author} {\bibfnamefont {Y.}~\bibnamefont {Oreg}}, \ and\ \bibinfo {author}
  {\bibfnamefont {D.}~\bibnamefont {Goldhaber-Gordon}},\ }\href
  {https://www.nature.com/articles/nature05556} {\bibfield  {journal} {\bibinfo
   {journal} {Nature}\ }\textbf {\bibinfo {volume} {446}},\ \bibinfo {pages}
  {167} (\bibinfo {year} {2007})}\BibitemShut {NoStop}%
\bibitem [{\citenamefont {Bulla}\ \emph {et~al.}(2008)\citenamefont {Bulla},
  \citenamefont {Costi},\ and\ \citenamefont {Pruschke}}]{Bulla2008}%
  \BibitemOpen
  \bibfield  {author} {\bibinfo {author} {\bibfnamefont {R.}~\bibnamefont
  {Bulla}}, \bibinfo {author} {\bibfnamefont {T.~A.}\ \bibnamefont {Costi}}, \
  and\ \bibinfo {author} {\bibfnamefont {T.}~\bibnamefont {Pruschke}},\ }\href
  {\doibase 10.1103/RevModPhys.80.395} {\bibfield  {journal} {\bibinfo
  {journal} {Rev. Mod. Phys.}\ }\textbf {\bibinfo {volume} {80}},\ \bibinfo
  {pages} {395} (\bibinfo {year} {2008})}\BibitemShut {NoStop}%
\bibitem [{\citenamefont {Mitchell}\ and\ \citenamefont
  {Bulla}(2015)}]{Mitchell2015}%
  \BibitemOpen
  \bibfield  {author} {\bibinfo {author} {\bibfnamefont {A.~K.}\ \bibnamefont
  {Mitchell}}\ and\ \bibinfo {author} {\bibfnamefont {R.}~\bibnamefont
  {Bulla}},\ }\href {\doibase 10.1103/PhysRevB.92.155101} {\bibfield  {journal}
  {\bibinfo  {journal} {Phys. Rev. B}\ }\textbf {\bibinfo {volume} {92}},\
  \bibinfo {pages} {155101} (\bibinfo {year} {2015})}\BibitemShut {NoStop}%
\bibitem [{\citenamefont {Metzner}\ \emph {et~al.}(2012)\citenamefont
  {Metzner}, \citenamefont {Salmhofer}, \citenamefont {Honerkamp},
  \citenamefont {Meden},\ and\ \citenamefont {Sch\"onhammer}}]{Metzner2012}%
  \BibitemOpen
  \bibfield  {author} {\bibinfo {author} {\bibfnamefont {W.}~\bibnamefont
  {Metzner}}, \bibinfo {author} {\bibfnamefont {M.}~\bibnamefont {Salmhofer}},
  \bibinfo {author} {\bibfnamefont {C.}~\bibnamefont {Honerkamp}}, \bibinfo
  {author} {\bibfnamefont {V.}~\bibnamefont {Meden}}, \ and\ \bibinfo {author}
  {\bibfnamefont {K.}~\bibnamefont {Sch\"onhammer}},\ }\href {\doibase
  10.1103/RevModPhys.84.299} {\bibfield  {journal} {\bibinfo  {journal} {Rev.
  Mod. Phys.}\ }\textbf {\bibinfo {volume} {84}},\ \bibinfo {pages} {299}
  (\bibinfo {year} {2012})}\BibitemShut {NoStop}%
\bibitem [{\citenamefont {Wingreen}\ and\ \citenamefont
  {Meir}(1994)}]{Wingreen1994}%
  \BibitemOpen
  \bibfield  {author} {\bibinfo {author} {\bibfnamefont {N.~S.}\ \bibnamefont
  {Wingreen}}\ and\ \bibinfo {author} {\bibfnamefont {Y.}~\bibnamefont
  {Meir}},\ }\href {\doibase 10.1103/PhysRevB.49.11040} {\bibfield  {journal}
  {\bibinfo  {journal} {Phys. Rev. B}\ }\textbf {\bibinfo {volume} {49}},\
  \bibinfo {pages} {11040} (\bibinfo {year} {1994})}\BibitemShut {NoStop}%
\bibitem [{\citenamefont {Horv\'ath}\ \emph {et~al.}(2011)\citenamefont
  {Horv\'ath}, \citenamefont {Lazarovits},\ and\ \citenamefont
  {Zar\'and}}]{Horvath2011}%
  \BibitemOpen
  \bibfield  {author} {\bibinfo {author} {\bibfnamefont {B.}~\bibnamefont
  {Horv\'ath}}, \bibinfo {author} {\bibfnamefont {B.}~\bibnamefont
  {Lazarovits}}, \ and\ \bibinfo {author} {\bibfnamefont {G.}~\bibnamefont
  {Zar\'and}},\ }\href {\doibase 10.1103/PhysRevB.84.205117} {\bibfield
  {journal} {\bibinfo  {journal} {Phys. Rev. B}\ }\textbf {\bibinfo {volume}
  {84}},\ \bibinfo {pages} {205117} (\bibinfo {year} {2011})}\BibitemShut
  {NoStop}%
\bibitem [{\citenamefont {Bickers}\ \emph {et~al.}(1989)\citenamefont
  {Bickers}, \citenamefont {Scalapino},\ and\ \citenamefont
  {White}}]{Bickers1989}%
  \BibitemOpen
  \bibfield  {author} {\bibinfo {author} {\bibfnamefont {N.~E.}\ \bibnamefont
  {Bickers}}, \bibinfo {author} {\bibfnamefont {D.~J.}\ \bibnamefont
  {Scalapino}}, \ and\ \bibinfo {author} {\bibfnamefont {S.~R.}\ \bibnamefont
  {White}},\ }\href {\doibase 10.1103/PhysRevLett.62.961} {\bibfield  {journal}
  {\bibinfo  {journal} {Phys. Rev. Lett.}\ }\textbf {\bibinfo {volume} {62}},\
  \bibinfo {pages} {961} (\bibinfo {year} {1989})}\BibitemShut {NoStop}%
\bibitem [{\citenamefont {Bickers}\ and\ \citenamefont
  {White}(1991)}]{Bickers1991}%
  \BibitemOpen
  \bibfield  {author} {\bibinfo {author} {\bibfnamefont {N.~E.}\ \bibnamefont
  {Bickers}}\ and\ \bibinfo {author} {\bibfnamefont {S.~R.}\ \bibnamefont
  {White}},\ }\href {\doibase 10.1103/PhysRevB.43.8044} {\bibfield  {journal}
  {\bibinfo  {journal} {Phys. Rev. B}\ }\textbf {\bibinfo {volume} {43}},\
  \bibinfo {pages} {8044} (\bibinfo {year} {1991})}\BibitemShut {NoStop}%
\bibitem [{\citenamefont {Gull}\ \emph {et~al.}(2011)\citenamefont {Gull},
  \citenamefont {Millis}, \citenamefont {Lichtenstein}, \citenamefont
  {Rubtsov}, \citenamefont {Troyer},\ and\ \citenamefont {Werner}}]{Gull2011}%
  \BibitemOpen
  \bibfield  {author} {\bibinfo {author} {\bibfnamefont {E.}~\bibnamefont
  {Gull}}, \bibinfo {author} {\bibfnamefont {A.~J.}\ \bibnamefont {Millis}},
  \bibinfo {author} {\bibfnamefont {A.~I.}\ \bibnamefont {Lichtenstein}},
  \bibinfo {author} {\bibfnamefont {A.~N.}\ \bibnamefont {Rubtsov}}, \bibinfo
  {author} {\bibfnamefont {M.}~\bibnamefont {Troyer}}, \ and\ \bibinfo {author}
  {\bibfnamefont {P.}~\bibnamefont {Werner}},\ }\href {\doibase
  10.1103/RevModPhys.83.349} {\bibfield  {journal} {\bibinfo  {journal} {Rev.
  Mod. Phys.}\ }\textbf {\bibinfo {volume} {83}},\ \bibinfo {pages} {349}
  (\bibinfo {year} {2011})}\BibitemShut {NoStop}%
\bibitem [{\citenamefont {Baym}(1962)}]{Baym1962}%
  \BibitemOpen
  \bibfield  {author} {\bibinfo {author} {\bibfnamefont {G.}~\bibnamefont
  {Baym}},\ }\href {\doibase 10.1103/PhysRev.127.1391} {\bibfield  {journal}
  {\bibinfo  {journal} {Phys. Rev.}\ }\textbf {\bibinfo {volume} {127}},\
  \bibinfo {pages} {1391} (\bibinfo {year} {1962})}\BibitemShut {NoStop}%
\bibitem [{\citenamefont {Kadanoff}\ and\ \citenamefont
  {Martin}(1961)}]{Kadanoff1961}%
  \BibitemOpen
  \bibfield  {author} {\bibinfo {author} {\bibfnamefont {L.~P.}\ \bibnamefont
  {Kadanoff}}\ and\ \bibinfo {author} {\bibfnamefont {P.~C.}\ \bibnamefont
  {Martin}},\ }\href {\doibase 10.1103/PhysRev.124.670} {\bibfield  {journal}
  {\bibinfo  {journal} {Phys. Rev.}\ }\textbf {\bibinfo {volume} {124}},\
  \bibinfo {pages} {670} (\bibinfo {year} {1961})}\BibitemShut {NoStop}%
\bibitem [{\citenamefont {Cornwall}\ \emph {et~al.}(1974)\citenamefont
  {Cornwall}, \citenamefont {Jackiw},\ and\ \citenamefont
  {Tomboulis}}]{Cornwall1974}%
  \BibitemOpen
  \bibfield  {author} {\bibinfo {author} {\bibfnamefont {J.~M.}\ \bibnamefont
  {Cornwall}}, \bibinfo {author} {\bibfnamefont {R.}~\bibnamefont {Jackiw}}, \
  and\ \bibinfo {author} {\bibfnamefont {E.}~\bibnamefont {Tomboulis}},\ }\href
  {\doibase 10.1103/PhysRevD.10.2428} {\bibfield  {journal} {\bibinfo
  {journal} {Phys. Rev. D}\ }\textbf {\bibinfo {volume} {10}},\ \bibinfo
  {pages} {2428} (\bibinfo {year} {1974})}\BibitemShut {NoStop}%
\bibitem [{\citenamefont {Sexty}\ \emph {et~al.}(2011)\citenamefont {Sexty},
  \citenamefont {Gasenzer},\ and\ \citenamefont {Pawlowski}}]{Sexty2011}%
  \BibitemOpen
  \bibfield  {author} {\bibinfo {author} {\bibfnamefont {D.}~\bibnamefont
  {Sexty}}, \bibinfo {author} {\bibfnamefont {T.}~\bibnamefont {Gasenzer}}, \
  and\ \bibinfo {author} {\bibfnamefont {J.}~\bibnamefont {Pawlowski}},\ }\href
  {\doibase 10.1103/PhysRevB.83.165315} {\bibfield  {journal} {\bibinfo
  {journal} {Phys. Rev. B}\ }\textbf {\bibinfo {volume} {83}},\ \bibinfo
  {pages} {165315} (\bibinfo {year} {2011})}\BibitemShut {NoStop}%
\bibitem [{\citenamefont {Bock}\ \emph {et~al.}(2016)\citenamefont {Bock},
  \citenamefont {Liluashvili},\ and\ \citenamefont {Gasenzer}}]{Sebastian2016}%
  \BibitemOpen
  \bibfield  {author} {\bibinfo {author} {\bibfnamefont {S.}~\bibnamefont
  {Bock}}, \bibinfo {author} {\bibfnamefont {A.}~\bibnamefont {Liluashvili}}, \
  and\ \bibinfo {author} {\bibfnamefont {T.}~\bibnamefont {Gasenzer}},\ }\href
  {\doibase 10.1103/PhysRevB.94.045108} {\bibfield  {journal} {\bibinfo
  {journal} {Phys. Rev. B}\ }\textbf {\bibinfo {volume} {94}},\ \bibinfo
  {pages} {045108} (\bibinfo {year} {2016})}\BibitemShut {NoStop}%
\bibitem [{\citenamefont {\ifmmode~\check{Z}\else \v{Z}\fi{}itko}\ and\
  \citenamefont {Bon\ifmmode~\check{c}\else \v{c}\fi{}a}(2007)}]{Zitko2007}%
  \BibitemOpen
  \bibfield  {author} {\bibinfo {author} {\bibfnamefont {R.}~\bibnamefont
  {\ifmmode~\check{Z}\else \v{Z}\fi{}itko}}\ and\ \bibinfo {author}
  {\bibfnamefont {J.}~\bibnamefont {Bon\ifmmode~\check{c}\else \v{c}\fi{}a}},\
  }\href {\doibase 10.1103/PhysRevB.76.241305} {\bibfield  {journal} {\bibinfo
  {journal} {Phys. Rev. B}\ }\textbf {\bibinfo {volume} {76}},\ \bibinfo
  {pages} {241305} (\bibinfo {year} {2007})}\BibitemShut {NoStop}%
\bibitem [{\citenamefont {\ifmmode~\check{Z}\else \v{Z}\fi{}itko}\ \emph
  {et~al.}(2012)\citenamefont {\ifmmode~\check{Z}\else \v{Z}\fi{}itko},
  \citenamefont {Mravlje},\ and\ \citenamefont {Haule}}]{Zitko2012}%
  \BibitemOpen
  \bibfield  {author} {\bibinfo {author} {\bibfnamefont {R.}~\bibnamefont
  {\ifmmode~\check{Z}\else \v{Z}\fi{}itko}}, \bibinfo {author} {\bibfnamefont
  {J.}~\bibnamefont {Mravlje}}, \ and\ \bibinfo {author} {\bibfnamefont
  {K.}~\bibnamefont {Haule}},\ }\href {\doibase 10.1103/PhysRevLett.108.066602}
  {\bibfield  {journal} {\bibinfo  {journal} {Phys. Rev. Lett.}\ }\textbf
  {\bibinfo {volume} {108}},\ \bibinfo {pages} {066602} (\bibinfo {year}
  {2012})}\BibitemShut {NoStop}%
\bibitem [{\citenamefont {Ding}\ \emph {et~al.}(2009)\citenamefont {Ding},
  \citenamefont {Ye},\ and\ \citenamefont {Dong}}]{Ding2009}%
  \BibitemOpen
  \bibfield  {author} {\bibinfo {author} {\bibfnamefont {G.~H.}\ \bibnamefont
  {Ding}}, \bibinfo {author} {\bibfnamefont {F.}~\bibnamefont {Ye}}, \ and\
  \bibinfo {author} {\bibfnamefont {B.}~\bibnamefont {Dong}},\ }\href {\doibase
  10.1088/0953-8984/21/45/455303} {\bibfield  {journal} {\bibinfo  {journal}
  {J. Phys.: Condens. Matter}\ }\textbf {\bibinfo {volume} {21}},\ \bibinfo
  {pages} {455303} (\bibinfo {year} {2009})}\BibitemShut {NoStop}%
\bibitem [{\citenamefont {Ding}\ \emph {et~al.}(2005)\citenamefont {Ding},
  \citenamefont {Kim},\ and\ \citenamefont {Nahm}}]{Ding2005}%
  \BibitemOpen
  \bibfield  {author} {\bibinfo {author} {\bibfnamefont {G.~H.}\ \bibnamefont
  {Ding}}, \bibinfo {author} {\bibfnamefont {C.~K.}\ \bibnamefont {Kim}}, \
  and\ \bibinfo {author} {\bibfnamefont {K.}~\bibnamefont {Nahm}},\ }\href
  {\doibase 10.1103/PhysRevB.71.205313} {\bibfield  {journal} {\bibinfo
  {journal} {Phys. Rev. B}\ }\textbf {\bibinfo {volume} {71}},\ \bibinfo
  {pages} {205313} (\bibinfo {year} {2005})}\BibitemShut {NoStop}%
\bibitem [{\citenamefont {Wong}\ \emph {et~al.}(2012)\citenamefont {Wong},
  \citenamefont {Lane}, \citenamefont {Dias~da Silva}, \citenamefont
  {Ingersent}, \citenamefont {Sandler},\ and\ \citenamefont
  {Ulloa}}]{Wong2012}%
  \BibitemOpen
  \bibfield  {author} {\bibinfo {author} {\bibfnamefont {A.}~\bibnamefont
  {Wong}}, \bibinfo {author} {\bibfnamefont {W.~B.}\ \bibnamefont {Lane}},
  \bibinfo {author} {\bibfnamefont {L.~G. G.~V.}\ \bibnamefont {Dias~da
  Silva}}, \bibinfo {author} {\bibfnamefont {K.}~\bibnamefont {Ingersent}},
  \bibinfo {author} {\bibfnamefont {N.}~\bibnamefont {Sandler}}, \ and\
  \bibinfo {author} {\bibfnamefont {S.~E.}\ \bibnamefont {Ulloa}},\ }\href
  {\doibase 10.1103/PhysRevB.85.115316} {\bibfield  {journal} {\bibinfo
  {journal} {Phys. Rev. B}\ }\textbf {\bibinfo {volume} {85}},\ \bibinfo
  {pages} {115316} (\bibinfo {year} {2012})}\BibitemShut {NoStop}%
\bibitem [{\citenamefont {Posazhennikova}\ \emph {et~al.}(2007)\citenamefont
  {Posazhennikova}, \citenamefont {Bayani},\ and\ \citenamefont
  {Coleman}}]{Posazhennikova2007}%
  \BibitemOpen
  \bibfield  {author} {\bibinfo {author} {\bibfnamefont {A.}~\bibnamefont
  {Posazhennikova}}, \bibinfo {author} {\bibfnamefont {B.}~\bibnamefont
  {Bayani}}, \ and\ \bibinfo {author} {\bibfnamefont {P.}~\bibnamefont
  {Coleman}},\ }\href {\doibase 10.1103/PhysRevB.75.245329} {\bibfield
  {journal} {\bibinfo  {journal} {Phys. Rev. B}\ }\textbf {\bibinfo {volume}
  {75}},\ \bibinfo {pages} {245329} (\bibinfo {year} {2007})}\BibitemShut
  {NoStop}%
\bibitem [{\citenamefont {Pohjola}\ \emph {et~al.}(2001)\citenamefont
  {Pohjola}, \citenamefont {Schoeller},\ and\ \citenamefont
  {Sch\"on}}]{Pohjola2001}%
  \BibitemOpen
  \bibfield  {author} {\bibinfo {author} {\bibfnamefont {T.}~\bibnamefont
  {Pohjola}}, \bibinfo {author} {\bibfnamefont {H.}~\bibnamefont {Schoeller}},
  \ and\ \bibinfo {author} {\bibfnamefont {G.}~\bibnamefont {Sch\"on}},\ }\href
  {http://stacks.iop.org/0295-5075/54/i=2/a=241} {\bibfield  {journal}
  {\bibinfo  {journal} {EPL}\ }\textbf {\bibinfo {volume} {54}},\ \bibinfo
  {pages} {241} (\bibinfo {year} {2001})}\BibitemShut {NoStop}%
\bibitem [{\citenamefont {Kubo}\ \emph {et~al.}(2008)\citenamefont {Kubo},
  \citenamefont {Tokura},\ and\ \citenamefont {Tarucha}}]{Kubo2008}%
  \BibitemOpen
  \bibfield  {author} {\bibinfo {author} {\bibfnamefont {T.}~\bibnamefont
  {Kubo}}, \bibinfo {author} {\bibfnamefont {Y.}~\bibnamefont {Tokura}}, \ and\
  \bibinfo {author} {\bibfnamefont {S.}~\bibnamefont {Tarucha}},\ }\href
  {\doibase 10.1103/PhysRevB.77.041305} {\bibfield  {journal} {\bibinfo
  {journal} {Phys. Rev. B}\ }\textbf {\bibinfo {volume} {77}},\ \bibinfo
  {pages} {041305} (\bibinfo {year} {2008})}\BibitemShut {NoStop}%
\bibitem [{\citenamefont {Zar\'and}\ \emph {et~al.}(2006)\citenamefont
  {Zar\'and}, \citenamefont {Chung}, \citenamefont {Simon},\ and\ \citenamefont
  {Vojta}}]{Zarand2006}%
  \BibitemOpen
  \bibfield  {author} {\bibinfo {author} {\bibfnamefont {G.}~\bibnamefont
  {Zar\'and}}, \bibinfo {author} {\bibfnamefont {C.-H.}\ \bibnamefont {Chung}},
  \bibinfo {author} {\bibfnamefont {P.}~\bibnamefont {Simon}}, \ and\ \bibinfo
  {author} {\bibfnamefont {M.}~\bibnamefont {Vojta}},\ }\href {\doibase
  10.1103/PhysRevLett.97.166802} {\bibfield  {journal} {\bibinfo  {journal}
  {Phys. Rev. Lett.}\ }\textbf {\bibinfo {volume} {97}},\ \bibinfo {pages}
  {166802} (\bibinfo {year} {2006})}\BibitemShut {NoStop}%
\bibitem [{\citenamefont {Galpin}\ \emph {et~al.}(2005)\citenamefont {Galpin},
  \citenamefont {Logan},\ and\ \citenamefont {Krishnamurthy}}]{Galpin2005}%
  \BibitemOpen
  \bibfield  {author} {\bibinfo {author} {\bibfnamefont {M.~R.}\ \bibnamefont
  {Galpin}}, \bibinfo {author} {\bibfnamefont {D.~E.}\ \bibnamefont {Logan}}, \
  and\ \bibinfo {author} {\bibfnamefont {H.~R.}\ \bibnamefont
  {Krishnamurthy}},\ }\href {\doibase 10.1103/PhysRevLett.94.186406} {\bibfield
   {journal} {\bibinfo  {journal} {Phys. Rev. Lett.}\ }\textbf {\bibinfo
  {volume} {94}},\ \bibinfo {pages} {186406} (\bibinfo {year}
  {2005})}\BibitemShut {NoStop}%
\bibitem [{\citenamefont {Keller}\ \emph {et~al.}(2016)\citenamefont {Keller},
  \citenamefont {Lim}, \citenamefont {S\'anchez}, \citenamefont {L\'opez},
  \citenamefont {Amasha}, \citenamefont {Katine}, \citenamefont {Shtrikman},\
  and\ \citenamefont {Goldhaber-Gordon}}]{Keller2016}%
  \BibitemOpen
  \bibfield  {author} {\bibinfo {author} {\bibfnamefont {A.~J.}\ \bibnamefont
  {Keller}}, \bibinfo {author} {\bibfnamefont {J.~S.}\ \bibnamefont {Lim}},
  \bibinfo {author} {\bibfnamefont {D.}~\bibnamefont {S\'anchez}}, \bibinfo
  {author} {\bibfnamefont {R.}~\bibnamefont {L\'opez}}, \bibinfo {author}
  {\bibfnamefont {S.}~\bibnamefont {Amasha}}, \bibinfo {author} {\bibfnamefont
  {J.~A.}\ \bibnamefont {Katine}}, \bibinfo {author} {\bibfnamefont
  {H.}~\bibnamefont {Shtrikman}}, \ and\ \bibinfo {author} {\bibfnamefont
  {D.}~\bibnamefont {Goldhaber-Gordon}},\ }\href {\doibase
  10.1103/PhysRevLett.117.066602} {\bibfield  {journal} {\bibinfo  {journal}
  {Phys. Rev. Lett.}\ }\textbf {\bibinfo {volume} {117}},\ \bibinfo {pages}
  {066602} (\bibinfo {year} {2016})}\BibitemShut {NoStop}%
\bibitem [{\citenamefont {S\'anchez}\ \emph {et~al.}(2010)\citenamefont
  {S\'anchez}, \citenamefont {L\'opez}, \citenamefont {S\'anchez},\ and\
  \citenamefont {B\"uttiker}}]{Sanchez2010}%
  \BibitemOpen
  \bibfield  {author} {\bibinfo {author} {\bibfnamefont {R.}~\bibnamefont
  {S\'anchez}}, \bibinfo {author} {\bibfnamefont {R.}~\bibnamefont {L\'opez}},
  \bibinfo {author} {\bibfnamefont {D.}~\bibnamefont {S\'anchez}}, \ and\
  \bibinfo {author} {\bibfnamefont {M.}~\bibnamefont {B\"uttiker}},\ }\href
  {\doibase 10.1103/PhysRevLett.104.076801} {\bibfield  {journal} {\bibinfo
  {journal} {Phys. Rev. Lett.}\ }\textbf {\bibinfo {volume} {104}},\ \bibinfo
  {pages} {076801} (\bibinfo {year} {2010})}\BibitemShut {NoStop}%
\bibitem [{\citenamefont {Kaasbjerg}\ and\ \citenamefont
  {Jauho}(2016)}]{Kaasbjerg2016}%
  \BibitemOpen
  \bibfield  {author} {\bibinfo {author} {\bibfnamefont {K.}~\bibnamefont
  {Kaasbjerg}}\ and\ \bibinfo {author} {\bibfnamefont {A.-P.}\ \bibnamefont
  {Jauho}},\ }\href {\doibase 10.1103/PhysRevLett.116.196801} {\bibfield
  {journal} {\bibinfo  {journal} {Phys. Rev. Lett.}\ }\textbf {\bibinfo
  {volume} {116}},\ \bibinfo {pages} {196801} (\bibinfo {year}
  {2016})}\BibitemShut {NoStop}%
\bibitem [{\citenamefont {Ding}\ and\ \citenamefont {Dong}(2013)}]{Ding2013}%
  \BibitemOpen
  \bibfield  {author} {\bibinfo {author} {\bibfnamefont {G.~H.}\ \bibnamefont
  {Ding}}\ and\ \bibinfo {author} {\bibfnamefont {B.}~\bibnamefont {Dong}},\
  }\href {\doibase 10.1103/PhysRevB.87.235303} {\bibfield  {journal} {\bibinfo
  {journal} {Phys. Rev. B}\ }\textbf {\bibinfo {volume} {87}},\ \bibinfo
  {pages} {235303} (\bibinfo {year} {2013})}\BibitemShut {NoStop}%
\end{thebibliography}%
\end{document}